\documentclass[11pt,a4paper,preprintnumbers]{article}
\usepackage{jcappub}
\pdfoutput=1
\usepackage[english]{babel}
\usepackage{amsmath,amssymb,amsfonts, bm,bbm,slashed, subdepth}
\usepackage{graphicx}
\usepackage[sort&compress]{natbib}
\usepackage{xcolor}
\usepackage[normalem]{ulem}
\usepackage{hyperref}
\usepackage{cleveref}
\definecolor{red}{rgb}{1.0, 0, 0}
\usepackage{hyperref}
\usepackage{enumerate}
\usepackage{epsfig, subfigure}
\usepackage{setspace}
\usepackage{booktabs, tabularx}
\usepackage{units}
\usepackage{hyperref}

\newcommand{\be}{\begin{equation}}
\newcommand{\ee}{\end{equation}}
\newcommand{\ba}{\begin{array}}
\newcommand{\ea}{\end{array}}
\newcommand{\bea}{\begin{eqnarray}}
\newcommand{\eea}{\end{eqnarray}}
\newcommand{\balg}{\begin{align}}
\newcommand{\ealg}{\end{align}}
\newcommand{\bit}{\begin{itemize}}
\newcommand{\eit}{\end{itemize}}
\newcommand{\trm}[1]{\textrm{#1}}

\newcommand{\Mpc}{\trm{\Mpc}}
\newcommand{\yr}{\trm{\yr}}
\newcommand{\eV}{\trm{\eV}}

\begin{document}

\title{Self-induced temporal instability from a neutrino antenna}

\author[a,b]{Francesco Capozzi,}
\author[c]{Basudeb Dasgupta,}
\author[d,e]{Alessandro Mirizzi}
\affiliation[a]{Dipartimento di Fisica e Astronomia ``Galileo Galilei,''\\ Via  Marzolo 8, 35131 Padova, Italy.}
\affiliation[b]{
Istituto Nazionale di Fisica Nucleare - Sezione di Padova,\\
Via  Marzolo 8, 35131 Padova, Italy.}
\affiliation[c]{Tata Institute of Fundamental Research,\\
             Homi Bhabha Road, Mumbai, 400005, India.}
\affiliation[d]{Dipartimento Interateneo di Fisica ``Michelangelo Merlin,''\\ Via Amendola 173, 70126 Bari, Italy.}
\affiliation[e]{
Istituto Nazionale di Fisica Nucleare - Sezione di Bari,\\
Via Amendola 173, 70126 Bari, Italy.}
\emailAdd{francesco.capozzi@pd.infn.it, bdasgupta@theory.tifr.res.in, alessandro.mirizzi@ba.infn.it}

\abstract{
It has been recently shown  that the flavor composition of a self-interacting neutrino gas can spontaneously acquire a time-dependent pulsating component during its  flavor evolution.  In this work, we perform a more detailed study of this effect in a model where neutrinos are assumed to be emitted in a two-dimensional plane from an infinite line that acts as a neutrino antenna. We consider several examples with varying matter and neutrino densities and find that temporal instabilities with various frequencies are excited in a cascade. We compare the numerical calculations of the flavor evolution with the predictions of linearized stability analysis of the equations of motion. The results obtained with these two approaches are in good agreement in the linear regime, while a dramatic speed-up of the flavor conversions occurs in the non-linear regime due to the interactions among the different pulsating modes. We show that large flavor conversions can take place if some of the temporal modes are unstable for long enough, and that this can happen even if the matter and neutrino densities are changing, as long as they vary slowly.}

\proceeding{TIFR/TH/16-06}
\maketitle

\section{Introduction}

The early universe and core-collapse supernovae (SNe) are 
remarkable environments where the neutrino density is so high that it  
can produce  self-induced refractive effects, associated with the 
neutrino-neutrino interactions, that render the flavor 
evolution to be highly non-linear~\cite{Dolgov:2002ab,Wong:2002fa,Duan:2006an,Hannestad:2006nj,Fogli:2007bk,Dasgupta:2007ws,Akhmedov:2016gzx}.
In the past decade, it has been realized that this effect can dramatically affect the flavor conversions 
in the deepest SN regions, producing collective oscillations of the dense neutrino gas, which could lead to
observable signatures on the  burst, e.g., spectral splits~\cite{Raffelt:2007cb,Duan:2007bt,Dasgupta:2009mg,Dasgupta:2010cd} or 
 flavor equilibration induced by fast conversions close to the neutrino-sphere~\cite{Sawyer:2008zs,Sawyer:2015dsa,Chakraborty:2016lct}.
The characterization of these effects represents a formidable challenge, considering the many layers
of complexity that have been identified in this problem (see~\cite{Duan:2010bg,Mirizzi:2015eza,Chakraborty:2016yeg} for  recent reviews).

One of the recent surprises in this context is that self-interacting neutrinos can {\emph{spontaneously break}} the space-time symmetries inherent to the system and its evolution because small deviations from these symmetries, arising from initial conditions, can be dramatically amplified during the subsequent flavor evolution~\cite{Raffelt:2013rqa,Raffelt:2013isa,Mirizzi:2013rla,Mirizzi:2013wda,Duan:2013kba,Mangano:2014zda,Duan:2014gfa,Mirizzi:2015fva,%
Mirizzi:2015hwa,Chakraborty:2015tfa,Duan:2015cqa,Abbar:2015mca,Abbar:2015fwa,Dasgupta:2015iia}.
In order to show how deviations from
the space-time symmetries of a system would affect the flavor evolution, a simple two-dimensional model, the so-called \emph{line model}, has been proposed~\cite{Duan:2014gfa}. In this case it is assumed that neutrinos are emitted in a two-dimensional plane
from an infinite line. By means
of a stability analysis of the linearized equations of motion~\cite{Duan:2014gfa}, it has been shown that if one
perturbs the initial symmetries of the flavor content along the boundary, then self-induced oscillations can spontaneously break the spatial symmetries. 
In particular, new flavor instabilities   can develop also at large
neutrino densities, where oscillations would have been otherwise suppressed due to
synchronization. In~\cite{Mirizzi:2015fva} a numerical study of the flavor evolution has been performed for this case. Subsequently, in~\cite{Chakraborty:2015tfa} it has been shown using a linear stability analysis that the presence of
a large matter term can suppress flavor conversions in both the homogeneous and the non-homogeneous case, due to  ``multi-angle'' matter effects~\cite{EstebanPretel:2008ni}.
However, the issue can be more subtle. Indeed, if one takes the consequences of the spontaneous symmetry breaking seriously, one should allow also for \emph{spontaneous non-stationarity} in the flavor evolution. In this context, in~\cite{Dasgupta:2015iia} two of us have recently shown in the line-model that the presence of an unstable pulsating mode can lead to flavor conversions at high neutrino and matter densities, as the frequency of pulsation can undo the phase dispersion due to a large matter density. The line model behaves in this case as a neutrino antenna, propagating a pulsating neutrino signal.

In the present work we will study the development of the temporal instability, choosing several representative cases and comparing, for each of them, the results of the numerical solution of the equations of motion with analogous results from a linear stability analysis. In Sec.\,\ref{sec:sec2} we describe the features of the non-stationary and inhomogeneous flavor evolution. We discuss the equations of motion for time-dependent two-dimensional case. We show how it is possible to solve this problem by Fourier transforming these equations, obtaining a tower
of ordinary differential equations for the different Fourier
modes in space and time. We also present the linear stability analysis of these equations, with explicit algebraic equations for the eigenvalues for the line model.
In Sec.\,\ref{sec:sec3} we discuss the  results of our
study for different representative cases. We find a good agreement between the results obtained with this two approaches in the linear regime, while a dramatic
amplification of the flavor conversions occur in the non-linear case due to the interaction among the different pulsating modes, that get 
excited in a cascade.
Finally, in Sec.\,\ref{sec:sec4} we comment 
about future developments and we conclude.

\section{Setup of the flavor evolution }
\label{sec:sec2}

\subsection{Equations of motion}
We consider the situation that neutrinos and antineutrinos are emitted from a surface and subsequently free-stream, but with forward scatterings with other neutrinos and antineutrinos as well as the background matter.
The flavor evolution can be characterized in terms of the (anti)neutrino density matrix $\varrho_{E,{\bf v}}$, for each neutrino mode with energy $E$ and velocity ${\bf v}$. In this formalism the diagonal elements of the density matrix represents the occupation numbers, while the off-diagonal terms encode phase relations
that allow one to follow flavor oscillations.
 Then, the equation of motion (EOM) for such a stream of neutrinos is~\cite{Sigl:1992fn,McKellar:1992ja,Vlasenko:2013fja}
\begin{align}
i(\partial_t + {\bf v}\cdot{\nabla}_{\bf x})\varrho_{E,{\bf v}}&=[{\sf H}_{E,{\bf v}},\varrho_{E,{\bf v}}]\,.
\label{eq:eom}
\end{align}
Of course $\varrho$ varies with time and space, i.e., $\varrho=\varrho(t,{\bf x})$. This space-time dependence is always understood, as also for other medium-dependent quantities, e.g., the background net electron density $n_e(t,{\bf x})$. For notational clarity, hereafter we shall omit showing these dependencies wherever there is no scope for confusion.

The Hamiltonian for flavor evolution in the collisionless limit is
\begin{align}
{\sf H}_{E,{\bf v}}&={\sf H}_{\rm vac}+{\sf H}_{\rm mat}+{\sf H}_{\nu\nu}\,,
\end{align}
where, in the two flavors limit,
\begin{align}
{\sf H}_{\rm vac}&=\frac{1}{2E}{\sf U}_\theta\left(\begin{array}{cc}m_1^2&0\\0&m_2^2\end{array}\right){\sf U}_\theta^\dagger\\
{\sf H}_{\rm mat}&=\sqrt{2}G_F\left(\begin{array}{cc}n_e&0\\0&0\end{array}\right)\\
{\sf H}_{\nu\nu}&=\sqrt{2}G_F\int\,\frac{1}{(2\pi)^3}E'^2dE' d{\bf v}'(1-{\bf v}\cdot{\bf v'})\varrho_{E',{\bf v}'}\,\label{eq:hnunu}\,,
\end{align}
where ${\sf U_\theta}$ is the mixing matrix. We introduce  the mass-squared difference  $\Delta m^2= m_2^2-m_1^2$, which is positive for normal mass ordering and negative for inverted ordering. The integral over energy spans $-\infty$ to $+\infty$, with ``negative'' $E$ corresponding to  antineutrinos of energy $E$, i.e., $\bar{\varrho}_E=-\varrho_{-E}$. Note the overall minus sign, so that equations for neutrinos and antineutrinos are the same with the replacement $E\to-E$. The integral over ${\bf v}$  corresponds  to integrals over the two independent velocity components, that are typically chosen such that $\varrho_{E,{\bf v}}(t,{\bf x})$ remains constant in the absence of oscillations.

The vacuum oscillation frequency $\omega=|\Delta m^2|/(2E)$ is a more convenient energy label than $E$. Thus, one has $\omega>0$ for neutrinos and $\omega<0$ for antineutrinos. The density matrix can be written as
\begin{align}
\varrho_{\omega,{\bf v}} = \frac{\rm Tr\varrho_{\omega,{\bf v}}}{2}{\rm I}+\frac{\Phi_\nu}{2}\,{g}_{\omega,{\bf v}}\left(\begin{array}{cc}s_{\omega,{\bf v}}&S_{\omega,{\bf v}}\\S^*_{\omega,{\bf v}}&-s_{\omega,{\bf v}}\end{array}\right)\,,
\end{align}
where $s^2=1-|S|^2$, with $S=0$ when no flavor evolution has occurred. The function $\Phi_\nu$ is a suitable normalization of the 
emission spectrum $g_{\omega,{\bf v}}$, defined by
\begin{align}
\int_{-\infty}^{0}d\Gamma\,{\Phi_\nu g_{\omega,{\bf v}}}=-(\Phi_{\bar{\nu}_e}-\Phi_{\bar{\nu}_x})\,,
\end{align}
where $d\Gamma=d\omega\,d{\bf v}$ and $\Phi_{\bar\nu_{e,x}}$  are the unoscillated flavor-dependent neutrino number fluxes at $\bf x$. The neutrino-antineutrino asymmetry parameter $\epsilon$ is given by
\begin{align}
1+\epsilon=(\Phi_{{\nu}_e}-\Phi_{{\nu}_\mu})/(\Phi_{\bar{\nu}_e}-\Phi_{\bar{\nu}_\mu})\,.
\end{align}

The equations of motion [Eq.\,(\ref{eq:eom})] can then be written expanding all the quantities on
the Pauli matrices $\sigma$ as 
\begin{align}
(\partial_t + {\bf v}\cdot{\nabla}_{\bf x}){\bf P}_{\omega,{\bf v}}=\left[-\omega{\bf B} + \lambda{\bf L}+\mu\int d\Gamma'(1-{\bf v}\cdot{\bf v'}){\bf P}_{\omega',{\bf v}'}\right]\times{\bf P}_{\omega,{\bf v}}\,.
\label{eq:eom2fl}
\end{align}
where 
\begin{equation}
\lambda=\sqrt{2}G_F n_e  \,\
\end{equation}
is the matter potential and
\begin{equation}
\mu=\sqrt{2}G_F \Phi_\nu \,\ 
\end{equation}
is the $\nu$-$\nu$ potential.
 The vector ${\bf P}_{\omega,{\bf v}}=g_{\omega,{\bf v}}({\rm Re}\,S_{\omega,{\bf v}},-{\rm Im}\,S_{\omega,{\bf v}},s_{\omega,{\bf v}})^T$, constructed from the density matrix, is a so-called polarization vector that encapsulates the flavor composition, while ${\bf B}=(\sin2\theta,0,\cos2\theta)^T$ and ${\bf L}=(0,0,1)^T$ are the analogous vectors constructed from ${\sf H}_{\rm vac}$ and ${\sf H}_{\rm mat}$, respectively. In this paper we present our results for a normal neutrino mass ordering, i.e., $m_1<m_2$, which leads to the minus sign in front of $\omega{\bf B}$ in Eq.\,(\ref{eq:eom2fl}), and for the inverted ordering one simply flips that minus sign to get the equivalent equations.

The problem boils down to calculating ${\bf P}_{\omega,{\bf v}}(t,{\bf x})$, given their values at the source as a function of time. Typically this boundary condition is taken to be stationary in time and homogeneous over the source, which may naively suggest that the solution ought to respect these symmetries as well. However, such a solution is often unstable to small spatial and temporal fluctuations, and it is important to ascertain the role of spontaneous breaking of these spatial and temporal symmetries. We will study this in both the linear and nonlinear regime, and we introduce the set-up and the notation in the next sections.

\subsection{Evolution of Fourier modes in the line model}

\begin{figure}[!h]
\centering
\includegraphics[width=0.9\textwidth]{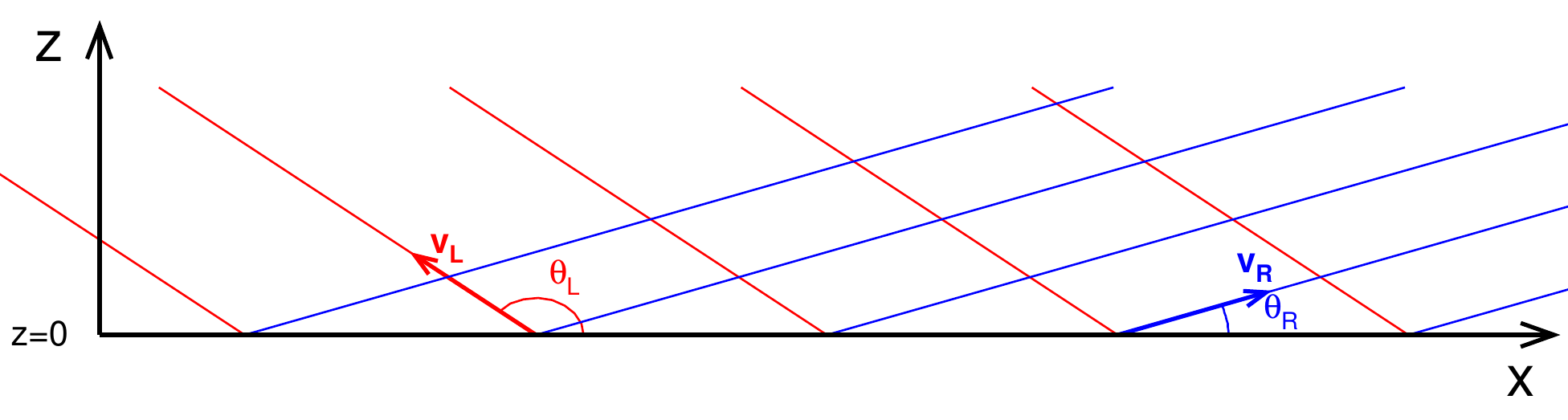}
\caption{Geometry of the line model used for the flavor evolution. Neutrinos and antineutrinos are emitted in two directions in a plane from every point of an infinite linear source (angles have been exaggerated for clarity). The flavor evolution is along $\hat{z}$, with fluctuations along $\hat{x}$ and time.
\label{line}}
\end{figure}

To study the role of these fluctuations concretely, we choose as toy-model of neutrino emission  the line model  where the growth of these instabilities is easily calculable.  The geometry of the model is shown in Fig.\,\ref{line}. Neutrinos and antineutrinos are emitted in the $x-z$ plane, in two directions ${\bf v}_L=\cos\vartheta_L\,\hat{x}+\sin\vartheta_L\,\hat{z}$ and ${\bf v}_R=\cos\vartheta_R\,\hat{x}+\sin\vartheta_R\,\hat{z}$, as shown. The dynamics is confined to this plane, though one may also think of this as a 3 dimensional problem with rotational symmetry around the $x$-axis. We consider neutrinos and antineutrinos emitted with a single energy $E_0$, so that
 \begin{equation}
 \omega_0= \frac{|\Delta m^2|}{2E_0} \,\ .
 \end{equation}  
This fixes the frequency scale of flavor evolution.  Therefore, the spectrum is given by
\begin{align}
g_{\omega,{\bf v}}=\big[-\delta(\omega+\omega_0)+(1+\epsilon)\delta(\omega-\omega_0)\big]\times\big[\delta({\bf v}-{\bf v}_L)+\delta({\bf v}-{\bf v}_R)\big]\,,
\end{align}
which implies the normalization $\Phi_\nu=(\Phi_{\bar{\nu}_e}-\Phi_{\bar{\nu}_\mu})/2$. $\Phi_\nu$ and $\lambda$ are constant along $x$
 and in $t$ but vary along $z$. 
Away from the source, the neutrino flavor composition can depend on $z$, as well as $x$ and $t$. The boundary conditions on ${\bf P}_{\omega,{\bf v}}$ will have to be specified along both $x$ and $t$ at $z=0$, and break the corresponding space-time translation symmetries.
 
The flavor evolution can be studied using Fourier modes of the polarization vectors. Eq.\,(\ref{eq:eom2fl}) can be converted to a tower of ordinary differential equations by decomposing the polarization vectors into their constituent Fourier modes labeled by $p$ and $k$~\cite{Mangano:2014zda,Duan:2014gfa,Mirizzi:2015fva}. One writes
\begin{align}
{\bf P}_{\omega,{\bf v}}(t,x,z) = \sum_{p,k}\,e^{-i(pt+kx)}{\bf P}^{p,k}_{\omega,{\bf v}}(z)\,,
\end{align}
where $p,k$ are the temporal and spatial frequency modes of the polarization vector. Using this decomposition, one gets a tower of equations for the Fourier modes ${\bf P}^{p,k}_{\omega,{\bf v}}$,
\begin{align}
v_z\partial_z{\bf P}^{p,k}_{\omega,{\bf v}}=i(p+v_{x}k){\bf P}^{p,k}_{\omega,{\bf v}}-(\omega{\bf B} -\lambda{\bf L})\times {\bf P}^{p,k}_{\omega,{\bf v}}+\sum_{p',k',\omega',{\bf v}'}\,\mu_{v}\,{\bf P}^{p-p',k-k'}_{\omega',{\bf v}'}\times{\bf P}^{p',k'}_{\omega,{\bf v}}\,.
\label{eq:four}
\end{align}
For our model, we have the simplifications, $\mu_{v}=\mu(1-{\bf v}_L\cdot{\bf v}_R)$, $\omega'$ only picks up the modes at $\pm\omega_0$, and ${\bf v}={\bf v}_L$ or ${\bf v}_R$. The velocity components, $v_x$ and $v_z$, being different for the $L$ and $R$ modes, produce multi-angle effects in additional to the spatial and temporal symmetry breaking in our set-up.

For our numerical studies, we assume that  only $\nu_e$  and ${\overline \nu_e}$ are emitted, with a
factor of two excess of $\nu_e$ over ${\overline \nu_e}$, i.e., $ \epsilon = 1$, chosen to guarantee that flavor conversions do not take place in the homogeneous case for the adopted value of the neutrino density $\mu$ on the boundary. We take the $L$ and $R$ modes to have two different angles $\vartheta_R = 5 \pi/18 $ and $\vartheta_L = 7 \pi/9$.  In this way, we mimic the multi-angle matter suppression in the presence of a large matter term. We choose $\theta=10^{-3}$ and a normal mass ordering, i.e., $\Delta m^2 > 0$, but the result would be qualitatively similar for the inverted ordering, i.e., $\omega{\bf B} \to -\omega{\bf B}$. The overall frequency-scale is set by $\omega_0 = 1$.

\subsection{Linearized stability analysis using Fourier modes}
The complete flavor evolution even in this rather simplified model is quite complicated, and it is useful to linearize the equations. To linear order in
 $S$, the EOMs [Eq.\,(\ref{eq:eom2fl})] simplify to \cite{Banerjee:2011fj,Chakraborty:2015tfa}
\begin{align}
i(\partial_t + {\bf v}\cdot{\nabla}_{\bf x})S_{\omega,{\bf v}}=(-\omega+\lambda+\epsilon\mu_v)S_{\omega,{\bf v}}
- \mu_v\int d\Gamma'g_{\omega',{\bf v}'}S_{\omega',{\bf v}'}\,.
\end{align}
One can again express $S_{\omega,{\bf v}}$ using its Fourier transform,
\begin{align}
S_{\omega,{\bf v}}(t,x,z)=\sum_{p,k}e^{-i(pt+{k}{x})}S^{p,{ k}}_{\omega,{\bf v}}(z)\,,
\end{align}
and take $S^{p,{ k}}_{\omega,{\bf v}}=Q^{p,{ k}}_{\omega,{\bf v}}e^{-i\Omega z}$. The eigenvalue equation for the Fourier amplitudes $Q^{p,{ k}}_{\omega,{\bf v}}$ is then given by
\begin{align}
\bigg(\frac{-\omega-{ k}{v}_x-p+{\lambda}+\epsilon\mu_v}{v_z}\bigg)Q^{p,{ k}}_{\omega,{\bf v}}-\frac{\mu_v}{v_z}
\int d\omega'\,d{\bf v'}\,
g_{\omega',{\bf v}'}
Q^{p,{k}}_{\omega',{\bf v}'}=\Omega Q^{p,{k}}_{\omega,{\bf v}}\,.
\label{eq:stab}
\end{align}
In the model we consider, there are only four neutrino momentum modes labeled by $\omega=\pm\omega_0$ and ${\bf v}={\bf v}_{L}$ and ${\bf v}_{R}$, and the above equation becomes an eigenvalue equation for the four modes, with 4 eigenvalues $\Omega_{i=1,2,3,4}$, as a function of $p$ and $k$. These eigenvalues are given by the equation
\begin{align}
{\rm Det}
\begin{pmatrix}
\omega_{+L}+\Omega&0&(1+\epsilon)\mu_{L}&-\mu_{{L}}\\
0&\omega_{{-L}}+\Omega&(1+\epsilon)\mu_{L}&-\mu_{{L}}\\
(1+\epsilon)\mu_{R}&-\mu_{{R}}&\omega_{+R}+\Omega&0\\
(1+\epsilon)\mu_{R}&-\mu_{{R}}&0&\omega_{{-R}}+\Omega
\end{pmatrix}=0\,,
\label{eq:eigv}
\end{align}
where,
\begin{align}
\omega_{\pm (L,R)}&=(\pm\omega_0+kv_{x,(L,R)}+p-\lambda-\epsilon\mu_v)/v_{z,(L,R)}\\
\mu_{(L,R)}&=\mu_v/v_{z,(L,R)}\,.
\end{align}
Unlike in the case of a symmetric set-up with $\vartheta_L=\vartheta_R$, there is no $k\to-k$ symmetry. This is however an artifact of explicitly breaking the $L\leftrightarrow R$ symmetry in order to mimic a multi-angle matter effect. These artifacts are avoided in a truly multi-angle scenario with $L\leftrightarrow R$ symmetric velocity distributions of many modes. Note, however, the characteristic equation is real and will only give real or complex-conjugate eigenvalues.

\begin{figure*}[!t]\centering
  \includegraphics[width=0.46\textwidth]{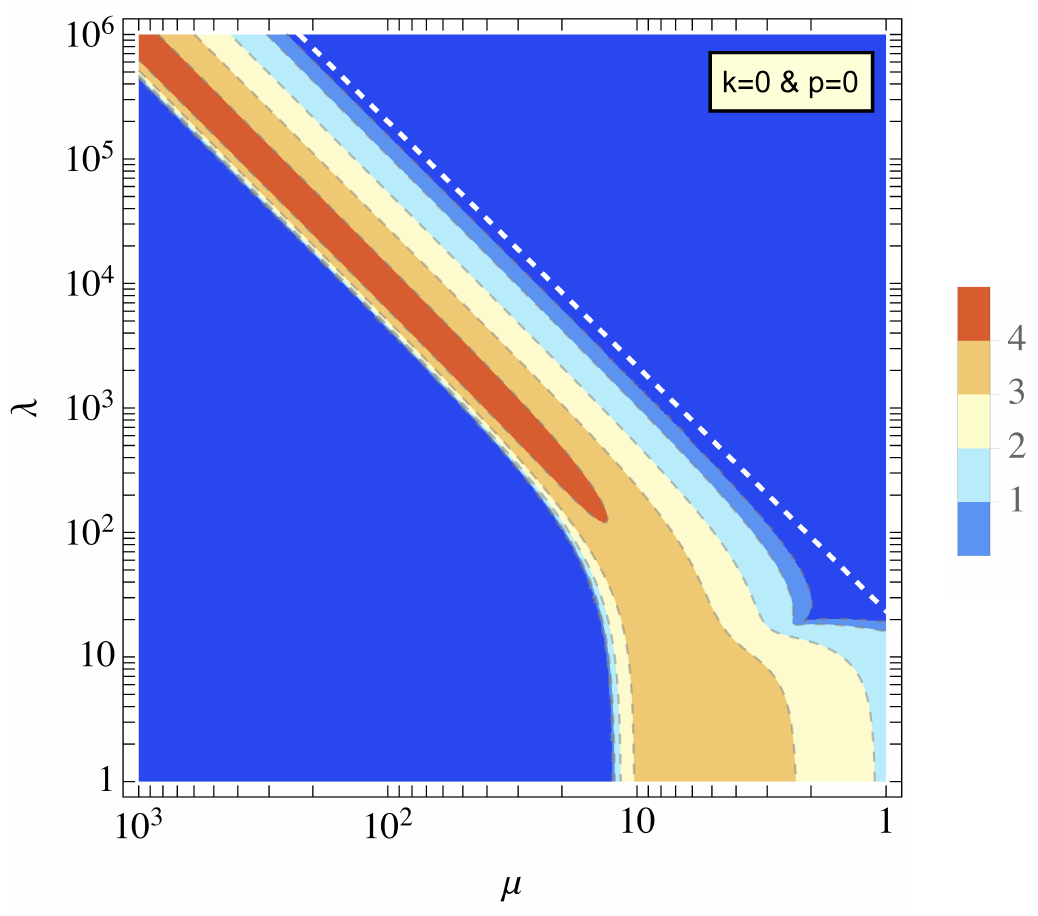}\hspace{2em}\includegraphics[width=0.46\columnwidth]{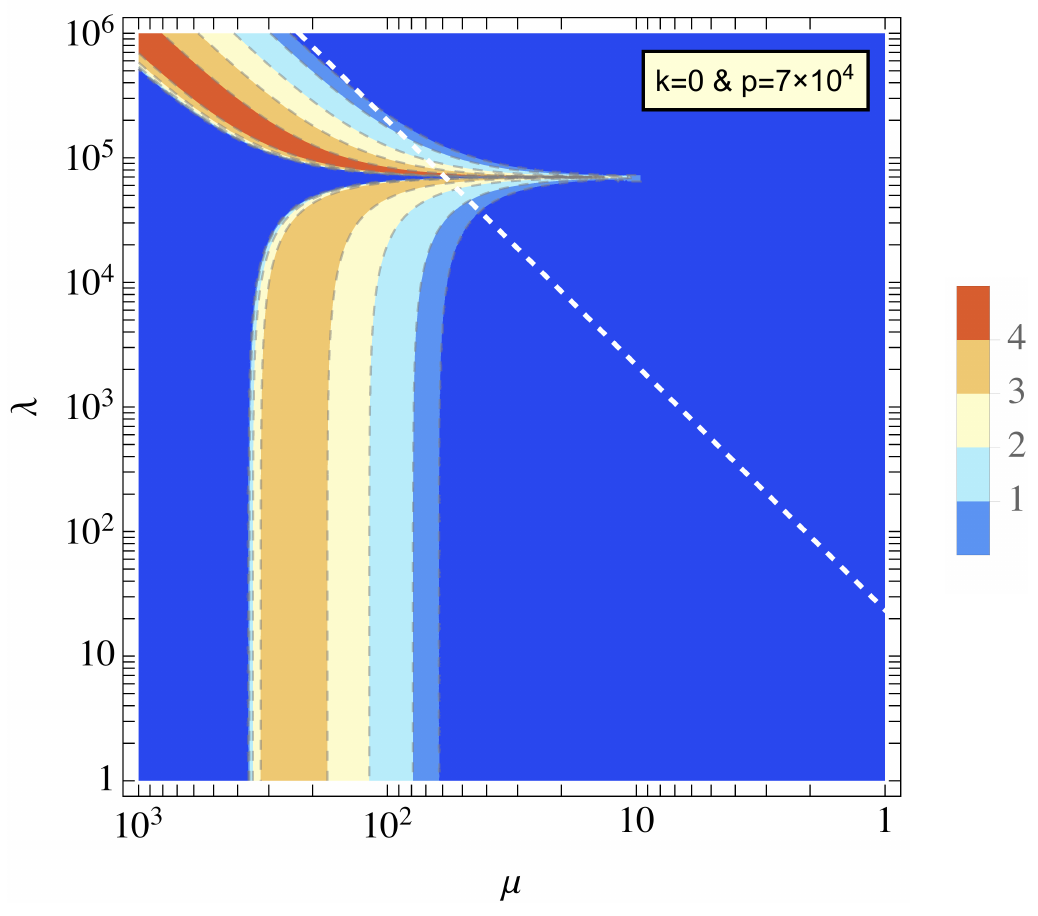}\\ \vspace{2em}  
  \includegraphics[width=0.46\textwidth]{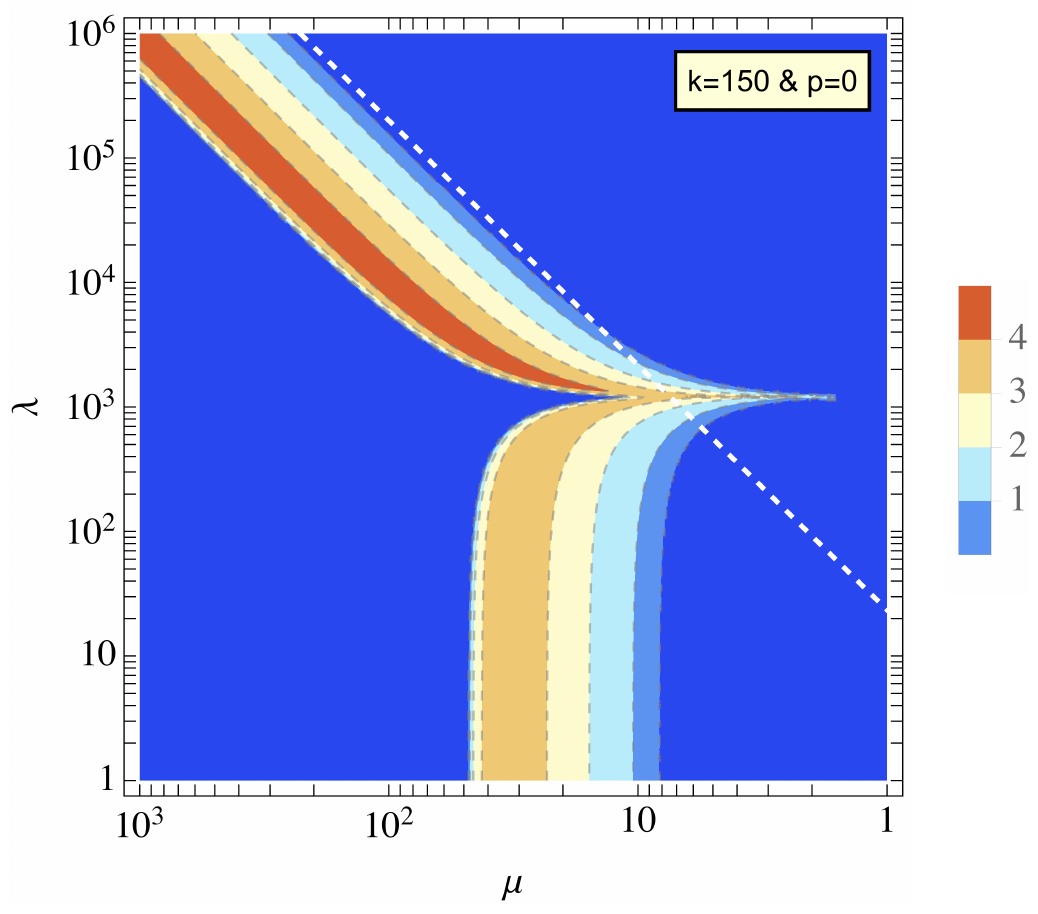}\hspace{2em}\includegraphics[width=0.46\columnwidth]{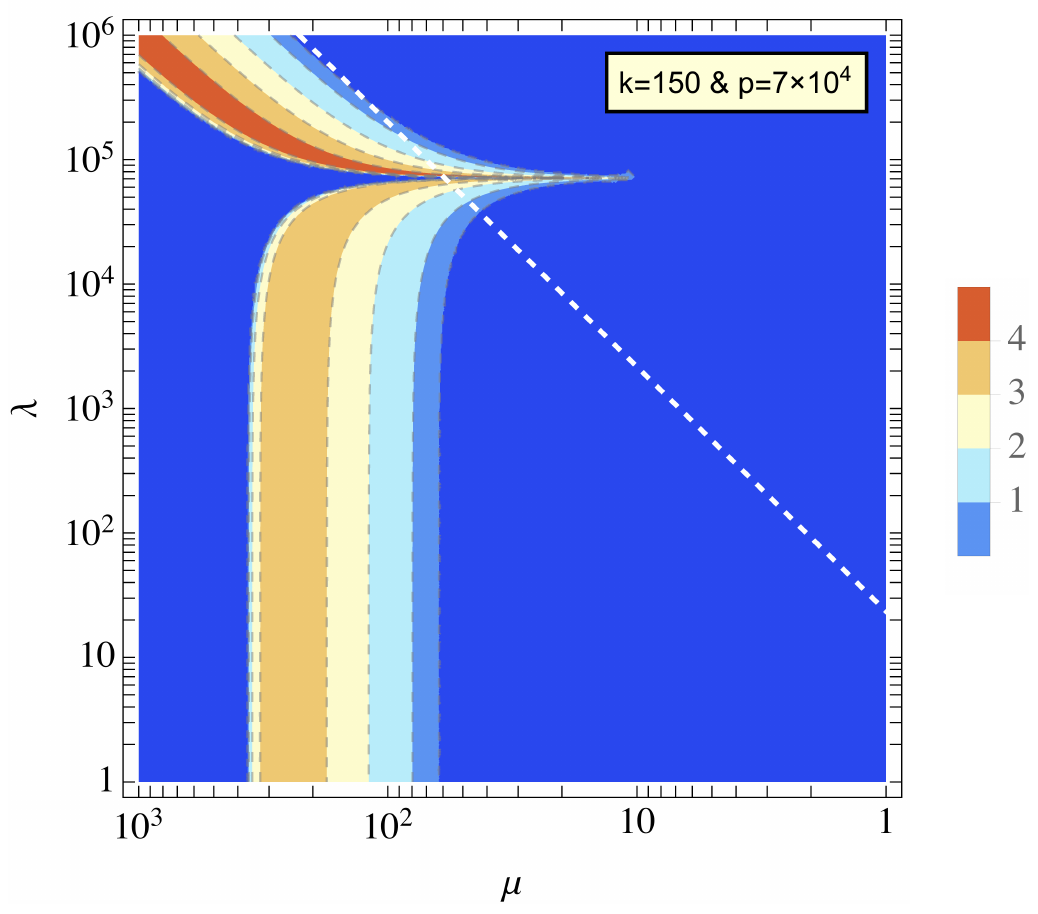}\\
  \caption{Footprints of instability for various values of  spatial and temporal frequency modes, $k$ and $p$ respectively, on the plane of the matter potential $\lambda$ and the neutrino potential $\mu$, all in units of $\omega_0$. The warmer colors represent larger magnitudes of the imaginary part of the eigenvalues given by Eq.\,(\ref{eq:eigv}). The dark blue color signifies that all eigenvalues are real. The dashed white line represents a correlation between $\mu$ and $\lambda$, e.g., in a model where both are varying with $z$ (see Sec.\,\ref{sec:3.2}). Note that while there are no instabilities for the model shown here as long as $k=0$ and $p=0$ (top left panel), instabilities appear when either $k\neq0$ or $p\neq0$ (all other panels).}
  \label{fig0}
\end{figure*}

The eigenvalues for this model can be calculated in a closed form, but these general expressions are not particularly illuminating. So we numerically evaluate these eigenvalues, and check if any of the eigenvalues have a large imaginary part that would signal rapid flavor conversions. Fig.\,\ref{fig0}, shows the largest imaginary part of any of the eigenvalues $\Omega$, for a range of $\lambda$ and $\mu$ and for specific values of $p$ and $k$. We see that for certain large values of $\lambda$ and $\mu$, for which there were no instabilities in the stationary and homogeneous scenario (top-left panel), become unstable to a pulsating mode with large $p$ (right panels) as the ``nose-like'' feature extends into these new regions. For a small number of modes as we have chosen here, a similar effect is produced by a choosing a large $k$ (bottom-left panel) but this cancellation becomes very fine-tuned when more velocity modes are included.

One important observation concerns the interplay of matter and temporal instability, i.e., $\lambda$ and $p$. A crucial feature of the temporal instabilities is that the eigenvalues depend on $\lambda$ only through the combination $\lambda+\epsilon\mu_v-p$, and there is always a $p$ that can remove the effect of matter for all modes, allowing an instability for  arbitrary $\lambda$. The temporal instability effectively undoes the phase dispersion introduced by matter and instabilities can grow unhindered by matter effects. This is seen in the right panels of Fig.\,\ref{fig0}. Such an effect is possible also for stationary but spatially inhomogeneous modes with $k\neq0$ (bottom-left panel in Fig.\,\ref{fig0}), but genuinely multi-angle matter effects with many modes will not be completely suppressed by them, unlike what we see here. This interplay remains true at a nonlinear level, which is best seen using Eq.\,(\ref{eq:four}). If one looks at a particular combination of the transverse components of the polarization vector, i.e., ${T}^{p,k}_{\omega,{\bf v}}\equiv({\bf P}^{p,k}_{\omega,{\bf v}})^{(1)}- i({\bf P}^{p,k}_{\omega,{\bf v}})^{(2)}$, its equations of motion look like 
\begin{equation}
v_z\partial_z {T}^{p,k}_{\omega,{\bf v}}= i(\lambda-p){T}^{p,k}_{\omega,{\bf v}}+\ldots\,,
\end{equation}
where the ellipsis refers to terms not depending on $\lambda$ or $p$. The amount of flavour conversion is determined by $\partial_z|{T}^{p,k}_{\omega,{\bf v}}|$, which clearly depends on $\lambda$ only via $\lambda-p$. Thus at a fully nonlinear level, a temporal pulsation $p$ counteracts the effect of matter density $\lambda$.

\section{Numerical examples}
\label{sec:sec3}

In this section we present the flavor evolution for the model described in the last section. We calculate the fully nonlinear evolution of the polarization vectors given by Eq.\,(\ref{eq:four}), and compare it to the predictions of linearized stability analysis given by Eq.\,(\ref{eq:eigv}). The off-diagonal component of the density matrix $\varrho_{\omega,{\bf v}}$, is proportional to $g_{\omega,{\bf v}}S_{\omega,{\bf v}}^{p,k}={T}^{p,k}_{\omega,{\bf v}}$. In our numerical calculations, we take $|{T}_{\omega,{\bf v}}^{0,0}|=10^{-7}$ for the $p=0, k=0$ mode, and all components of all other Fourier modes of the polarization vectors initially get seeded by numerical noise related to the accuracy of the numerical solution, i.e., ${\mathcal O}(10^{-12})$. Thus, the initial conditions are $|S_{\omega,{\bf v}}^{p,k}|\simeq\sqrt{2}\times 10^{-12}$ for all nonzero $p$. Their growth is predicted using the eigenvalues found through the linear analysis. The notation $\varrho^{e\mu}_{n_p,n_k}$ is used to refer to both ${T}^{p,k}_{\omega,{\bf v}}$ and $g_{\omega,{\bf v}}S_{\omega,{\bf v}}^{p,k}$, and measures the degree of flavor conversion in a given Fourier mode\footnote{The careful reader will notice that we drop the $(\omega, {\bf v})$ index here. Indeed, we take the $(+\omega_0,{\bf v}_R)$ mode from nonlinear calculations and compare it with the fastest mode in linear theory. The different modes are all nonlinearly coupled, and we expect the growth rate to be dominated by the fastest mode.}.

In general, one should consider the flavor evolution
of different Fourier modes associated with both spatial and temporal inhomogeneities.
However, this would be numerically a rather challenging task. Therefore, we select a particular  mode in space, i.e., the homogeneous mode with $k=0$, and follow only the development 
of the Fourier modes  in the $n_p$ space, suppressing the $n_k$ index henceforth\footnote{We thank Georg Raffelt for pointing out that one can consistently solve for only the $k=0$ homogeneous Fourier mode while ignoring other modes. Considering any other $n_k$ requires tracking the cascade in $n_k$.}. In order to make $p$ dimensionless, it is expressed in multiples of the effective matter potential 
 \begin{equation}
  {\bar\lambda=\lambda+\epsilon\mu_v} \,\ ,
  \end{equation}
at $z=0$, i.e.,
 \begin{equation}
 p= n_p \times \frac{{\bar \lambda(z=0)}}{100} \,\ .
 \end{equation}
All other quantities are measured in units of $\omega_0$. We include 300 $p$ modes, but with the modes $n_p > 200$ being kept empty. This trick avoids ``spectral blocking'' that leads to a spurious rise of the Fourier coefficients at large $n_p$ due to truncation of the tower of equations~\cite{spectral}. 
 We observed this effect if we do not use this prescription. However, we checked that using different number of modes but taking
the last $1/3$ of them empty, it does not affect the behavior of the modes which are not close to the largest $n_p$.
In the following, the corresponding Fourier modes for the nonlinear computation and the linearized stability analysis calculation are shown and compared.
\subsection{Constant $\mu$ and declining $\lambda$}
We present our first set of results with
\begin{table*}[!h]
\centering
\begin{tabular}{lll}
Constant neutrino potential $\mu$	&=& $40$\,, \\
Varying matter density $\lambda$ 	&=& $\lambda_0 \times \textrm{exp}(-z/\tau_\lambda)$\,,\\
Initial matter density $\lambda_0$  &=& $4\times10^4$\,,\\
Scale height $\tau_\lambda$			&=& $\infty,\, 30,\, {\rm and},\,10$\,.
\end{tabular}
\end{table*}

\begin{figure}[!t]
\centering
\includegraphics[height=6cm, width=0.95\textwidth]{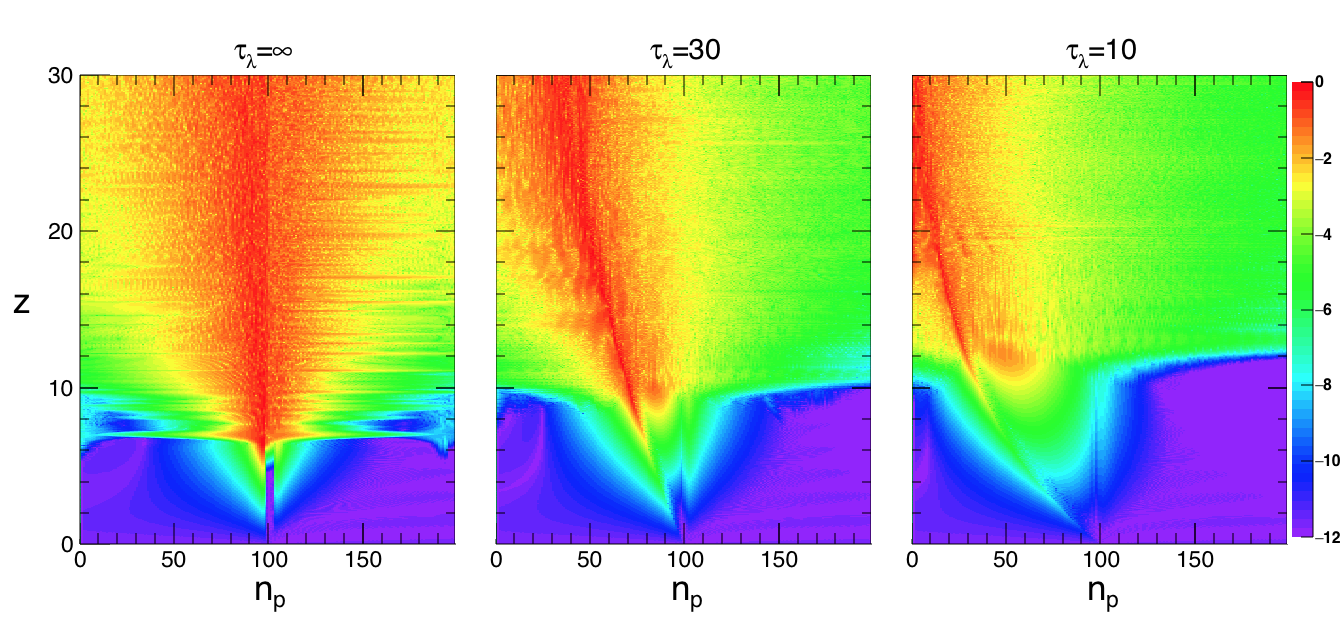}
\caption{Amplitudes of  the off-diagonal components of the different $n_p$ modes, $A^{e \mu}$,
 as a function of the distance from the neutrino source, $z$, and of the Fourier 
 mode with index $n_p$, for $\tau_\lambda=\infty$ (left panel),
30 (middle panel) and 10 (right panel) respectively. The closer is $A^{e \mu}$ to 
0 (red color in the plot), the stronger are the 
flavor conversions.
\label{fig2}}
\vspace{0.5cm}
\hspace{-0.8cm}
\centering
\includegraphics[width=0.95\textwidth,height=5.5cm]{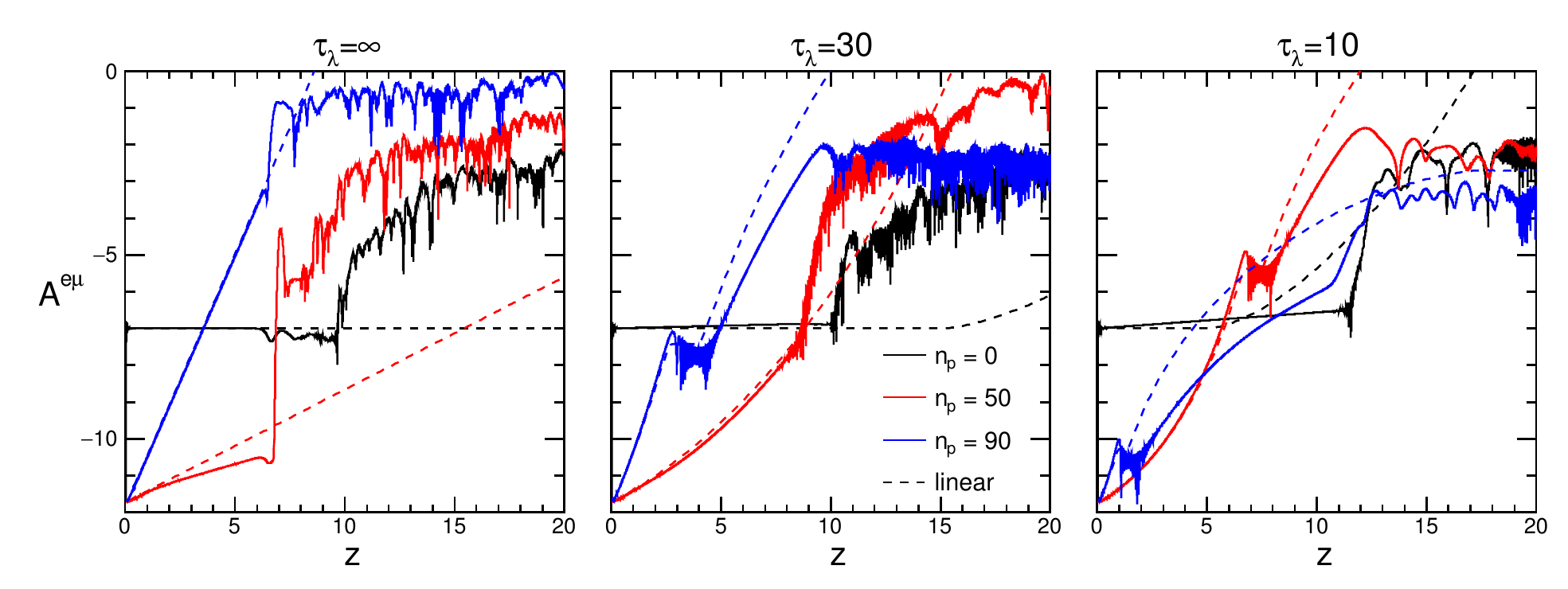}
\caption{Comparison linear vs nonlinear growth of $A^{e \mu}$
 for different Fourier modes
for $\tau_\lambda=\infty$ (left panel),
30 (middle panel) and 10 (right panel) respectively. 
\label{fig3}}
\end{figure}

In Fig.\,\ref{fig2}, the amplitudes of 
 the off-diagonal components of the different $n_p$ modes, 
\begin{equation}
A^{e \mu} (n_p,z) \equiv \textrm{log}_{10} |\varrho^{e \mu}_{n_p}(z)|\,,
\end{equation}
are shown at distance  $z$ from the emitting line for various Fourier modes $n_p$
for $\tau_\lambda=\infty$ (left panel),
30 (middle panel) and 10 (right panel) respectively. The key observations to be made from Fig.\,\ref{fig2} are
\begin{itemize}

\item The choice $\tau_\lambda=\infty$, corresponds to a constant matter density, $\lambda=\lambda_0$. Here, in agreement with the stability analysis, we find that the Fourier modes around $n_p \sim 100$  are excited first (at $z\simeq 2$). Indeed, for them the self-induced pulsation compensates the phase dispersion associated with the matter term ($p \simeq {\bar\lambda}$). 

At larger $z$ the temporal instability propagates in the Fourier space, where scales both larger and smaller than $n_p \sim 100$ start to get excited due to the non-linear interaction among the different modes [c.f. Eq.\,(\ref{eq:eom2fl})]. At $z=30$ all the modes with $n_p \gtrsim 50$ are sufficiently large to be nonlinear.

The finger-like feature just above $n_p\approx100$ arises due to the growth rate being zero for those modes. For those values of $p$ one finds only real eigenvalues.

\item For $\tau_\lambda=30$, the temporal instability starts at  $n_p \sim 100$. However, the finger-like feature just above $n_p\approx100$ now sweeps into lower $n_p$ due to the decline of $\lambda$, and these modes go off-resonance and their growth slows down. 

The cancellation of the matter suppression shifts to lower $n_p$ at larger $z$. As a result, the modes that are first to have $A^{e \mu} {\simeq -1}$ are $n_p \sim 80$ at $z\simeq 5$. Thereafter, the instability diffuses in the Fourier space.

With respect to the previous case of constant $\lambda$, lower $n_p$ are favored, because they are excited by both the non-linear effect and by the $p \simeq {\bar\lambda}$ condition that is satisfied at lower $p$.

\item Finally, in the case with $\tau_\lambda=10$, the rapid decline of $\lambda$ enhances the features of the previous case. In particular, the reduced adiabaticity of the evolution implies that the instability is reduced. Modes that satisfy the $p \simeq {\bar\lambda}$ condition go rapidly off-resonance due to the rapid change of $\lambda$. As a consequence, their further growth gets inhibited. 

The instability develops only at $z \gtrsim 10$ for $n_p \sim 50$. Modes with $n_p \lesssim 50$ grow till $A^{e \mu} {\simeq -1}$, when nonlinearity takes over, while larger $n_p$ modes are suppressed by the large $p$ that mimics the matter suppression.

\end{itemize}

In Fig.\,\ref{fig3}, we show that the above observations are justified by comparing the results of the nonlinear computation (continuous curves) with the predictions of linearized stability analysis (dashed curves) for the amplitudes  $A^{e \mu}$ of specific $n_p$ modes ($n_p=0,\, 50,\,{\rm and}\, 90$). The key points to be noted are

\begin{itemize}
\item In the case of $\tau_\lambda=\infty$, the mode at $n_p=90$ is the most unstable and 
grows by $\sim 10$ orders of magnitude till $z\simeq 6$. In the same range the mode at $n_p=50$ grows by only $2$ orders of magnitude, while the $n_p=0$ mode is stable. Till this distance from the source there is perfect agreement between the non-linear and the linear evolution for these three modes.

At larger $z$ when the mode with $n_p=90$ has $A^{e \mu} \simeq {-1}$. Its evolution becomes non-linear and produces significant deviations with respect to the linear case due to the interactions of different Fourier modes. In particular, the growth of the $n_p=90$ mode slows down with respect to the linear prediction and instead the mode at $n_p=50$ has a rapid increase of 5 orders of magnitude around $z=7$, which then continues to rise and reaches $A^{e \mu} \simeq {-2}$ at $z=20$. The mode at $n_p=0$ begins to  oscillate around its initial position at $z\gtrsim 7$ and starts growing at $z \gtrsim 10$. Its final value is only 5 orders of magnitude larger than its initial value, as the cascade in the Fourier space has not reached the $n_p=0$ mode in the considered $z$ range.

\item With $\tau_\lambda=30$, we realize that until $z\simeq 3$ the behaviour of the three modes is similar to the one found in the case of $\tau_\lambda=\infty$. In particular, the $n_p=90$ mode is still the most unstable, and grows by 4 orders of magnitude. However, this mode then undergoes rapid oscillations from $z=2$ to $z=4$ as $p$ is just larger than $\bar{\lambda}$, where the linear theory correctly predicts no growth (the dip feature here is the same as the finger in the Fig.\,\ref{fig2}). The agreement between linear and non-linear results is good till $z=4$ for the mode at $n_p=90$.

However, at larger $z$, as the matter potential becomes smaller than $p$, this mode grows less rapidly. As soon as the  $A^{e \mu}$ is affected by the rapid (non-adiabatic) oscillations owing to the dip-like feature in the growth rates, the linear solution does not reproduce this behavior and starts to deviate from the non-linear behavior. As a consequence at larger $z$ the linear solution overestimates the real growth.

On the other hand, the  mode $n_p=50$ steadily grows and  is enhanced by non-linear effects at $z\simeq 10$. For the mode $n_p=50$ the agreement is good till $z=10$, when the mode reaches $A^{e \mu} \gtrsim 10^{-2}$ and non-linearities begin to play an important role. Again, the growth of the $n_p=0$ mode is severely underestimated by linear theory, as it ignores the possibility nonlinear coupling of modes beyond $z=10$ that leads to an increase of 4 order of magnitudes by $z=20$.

\item  Finally, for $\tau_{\lambda}=10$ we see that due to the fast decline of $\lambda$, 
the growth of the mode $n_p=90$ is inhibited already at $z=1$, while the mode with $n_p=50$ steadily increases till  $z=6$. Then it oscillates (when $p$ is just larger than $\bar{\lambda}$) before resuming its growth at $z\gtrsim 7$. The agreement between the linear and non-linear result is good till $z=1$ for the mode at $n_p=90$, and for $z=5$ for the mode at $n_p=50$. In both the cases the discrepancy occurs when the mode deviates from a monotonic growth and starts to show these fast oscillations.

We see that unlike in the two previous cases, the mode with $n_p=0$ is not linearly stable
for $z>7$. In this case its linear growth is larger than  the non-linear one. In particular, at 
$z=12$ they disagree by 2 orders of magnitude. 
\end{itemize}

From these comparisons we conclude that the linear solution is a good approximation of the non-linear evolution only until nonlinearities remain small, so that the interaction among the different modes can be neglected, as evident from the  case with $\tau_\lambda=\infty$. Once nonlinearities kick-in, the rate of growth is severely underestimated by linear theory and the growth of flavor conversions appears to be qualitatively larger.  Furthermore, in the case of variable $\lambda$  the adiabaticity of the flavor evolution plays a significant role. Indeed, when the numerical solution shows non-adiabatic features resulting from resonances (rapid  oscillations in the dip-like feature), the deviations with respect to the linear approximation become significant.
\subsection{Declining $\mu$ and $\lambda$}
\label{sec:3.2}
As a mock-up for a more realistic SN-like scenario, it is interesting to consider situations in which both $\mu$ and $\lambda$ decline as a function of $z$. Fig.\,\ref{fig0} provides some guidance as to how $\lambda$ and $\mu$ should vary with $z$, so that there is no instability if $p$ and $k$ are zero but an instability develops when they are nonzero. Using this guidance, we consider that
$$\mu(z)=\mu_0 \times \textrm{exp}(-z/\tau_\mu)\,,$$
$$\lambda(z)=\lambda_0 \times \textrm{exp}(-z/\tau_\lambda)\,,$$
with the following parameters in units of $\omega_0$,
\begin{table*}[!h]
\centering
\begin{tabular}{lllll}
             && Model A     && Model B     \\
$\lambda_0$    && $7\times10^4$ && $1\times10^5$ \\
$\mu_0$        && 60          && 60          \\
$\tau_\lambda$ && 15          && 7.5         \\
$\tau_\mu$     && 30          && 15         
\end{tabular}
\end{table*}

The functions are chosen such that $\lambda$ and $\mu$ for model A varies as per the path shown as the dashed white line in Fig.\,\ref{fig0}. However, how fast the system evolves along the curve is determined by $\tau_\lambda$ and $\tau_\mu$. It is important to emphasize that while model B also follows almost the same path on the $\lambda-\mu$ plane, it passes through the instability much quicker than model A, owing to its smaller scale heights. Note that no instabilities would be encountered for either model if $k=0$ and $p=0$.

\begin{figure}[!t]
\centering
\includegraphics[width=0.92\textwidth]{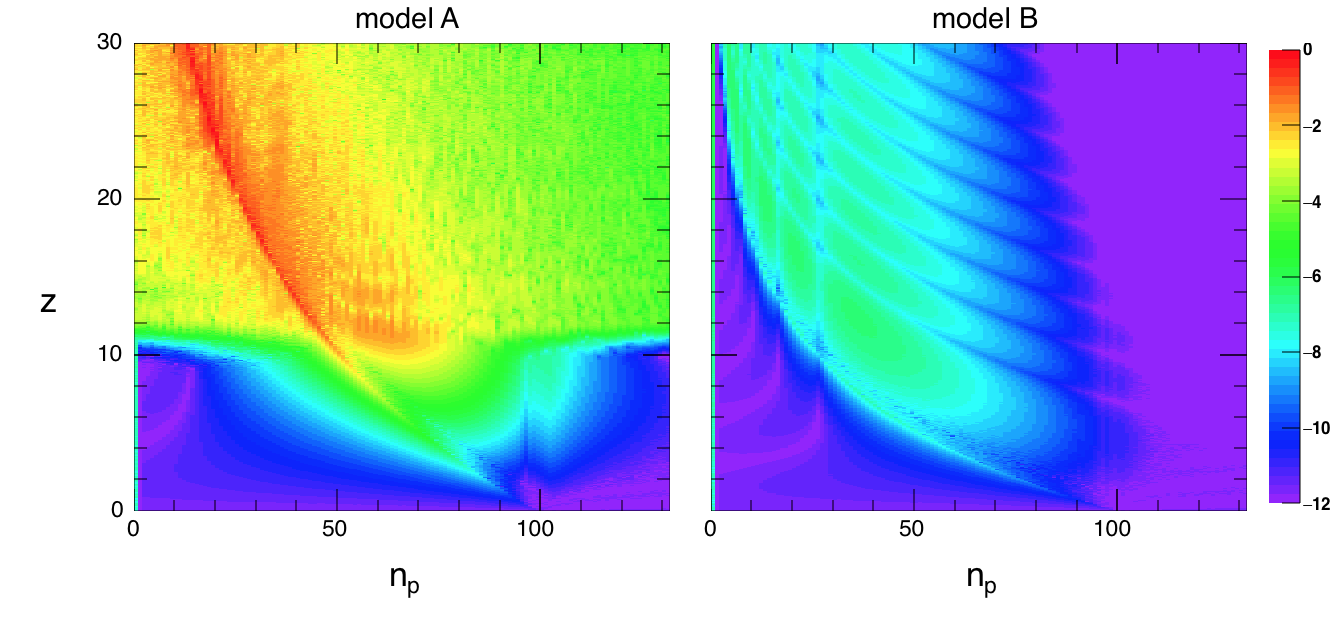}
\caption{Amplitudes of  the off-diagonal components of the different $n_p$ modes, $A^{e \mu}$,
 as a function of the distance from 
the neutrino
source $z$ and of the Fourier mode with index $n_p$, for the models A (left panel)
and B (right panel). See the text for details.
\label{fig4}}
\vspace{0.3cm}
\hspace{-0.8cm}
\centering
\includegraphics[width=0.87\textwidth]{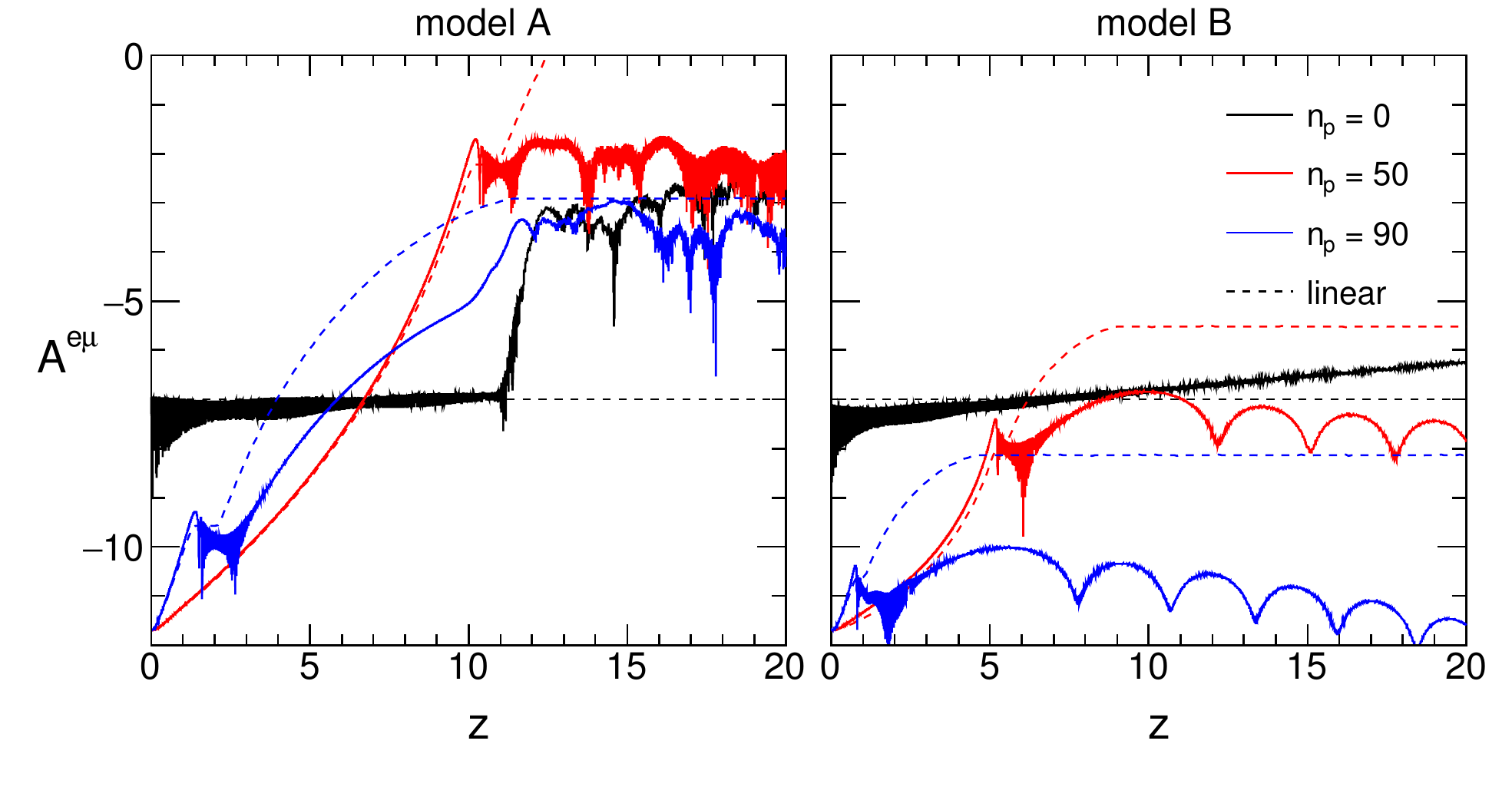}
\caption{Comparison linear vs nonlinear growth of $A^{e \mu}$
 for different Fourier modes of the models A (left panel), and B (right panel) respectively.
\label{fig5}}
\end{figure}

In Fig.\,\ref{fig4} the amplitudes of 
flavor conversions, $A^{e \mu}$,
are shown at distance $z$ for various Fourier modes $n_p$ for the models A (left)
and B (right). For the model A, we realize that the instability starts 
at $z\simeq 2$ from modes at $n_p \simeq 100$ and develops at smaller  
scales, reaching the mode at $n_p=0$ at $z=30$. On the other hand for model B,
due to the reduced adiabaticity in the flavor evolution the instability never
grows significantly.
This result tells that situations that present the same instability footprint, 
may have a rather different flavor evolution depending on net time spent in the unstable region and on the adiabaticity of evolution. If the profiles decline too fast the instability is inhibited.

Fig.\,\ref{fig5} is the analogue of Fig.\,\ref{fig3}, obtained by 
considering models A (left panel) and B (right panel).
The results here are similar to what is presented in the previous section in the case
of constant $\mu$. In particular, for model A we find that indeed the $p=0$ stationary flavor conversion mode can be nonlinearly excited by its coupling to the fastest $p\neq0$ modes by $z=11$, which would not be predicted in a purely linear theory. That aside, agreement between linear and non-linear evolution is good till the  non-linear evolution  presents non-adiabatic features (fast oscillations in the dip-like feature) which are not accounted in the linear theory.

\section{Conclusions}
\label{sec:sec4}

In our work we have studied the development of the temporal instability in self-induced 
flavor evolution in the line model. We have found that space and time inhomogeneities 
can excite flavor conversions at small spatial scales and can spontaneously generate a pulsating component in the flavor composition. 
In particular, the pulsation can compensate the phase dispersion associated with a 
large matter term that would otherwise suppress the flavor conversions.

We have considered several scenarios, with varying matter and neutrino densities, 
finding that the time instability propagates in the Fourier space exciting in 
cascade different Fourier modes. The development of this cascade crucially depends 
on the adiabaticity of the matter and neutrino potential. We have shown that the 
linearized solution of the equations of motion can be a valuable tool to find 
the growing modes and to predict the flavor evolution till non-linearities and 
non-adiabatic features become important. However, the linear theory does not 
account for the interaction among different Fourier modes that can enhance or reduce
the development of the pulsating modes.

This raises the possibility that a specific pulsating mode may 
experience large growth and eventually lead to flavor decoherence, especially if 
there is a slowly varying or relatively flat region of the matter density profile 
that extends for $\sim10^2\,$km, as sometimes found to be the case behind the 
shock in SN simulations (see, e.g.,~\cite{Tomas:2004gr,Fischer:2009af}), that allows long-term 
adiabatic growth of a single mode. Our model of course is too simple to gain a 
definite answer on the development 
of such effects in the flavor conversions on SN neutrinos. In particular, in SN 
neutrino context one has to take into account true multi-angle effects that would 
make the simulation of the non-linear flavor evolution extremely challenging. 
However, one crucial point that is illustrated by our study is that one may need 
to account for these details of the matter and neutrino profiles in a SN in order 
to get a definitive answer on the nature of neutrino flavor conversions in SNe.

\section*{Acknowledgments}
We thank Eligio Lisi and Georg Raffelt for many useful comments on the manuscript. The work of F.C. and A.M. is supported by the Italian Ministero dell'Istruzione, Universit\`a e Ricerca (MIUR) and Istituto Nazionale di Fisica Nucleare (INFN) through the ``Theoretical Astroparticle Physics'' projects.
F.C. also acknowledge support of the MIUR grant for the Research Projects of National Interest
PRIN 2012 No. 2012CPPYP7 Astroparticle Physics.

\end{document}